\documentstyle[12pt,epsf]{article}

\def\bpartial{\mbox{\boldmath$\partial$}}
\def\bnabla{\mbox{\boldmath$\nabla$}}

\def\bgamma{\mbox{\boldmath$\gamma$}}
\def\bpi{\mbox{\boldmath$\pi$}}

\def\bp{{\bf p}}
\def\br{{\bf r}}
\def\bv{{\bf v}}
\def\bz{{\bf z}}
\def\bx{{\bf x}}
\def\by{{\bf y}}
\def\bs{{\bf s}}

\def\bL{{\bf L}}

\def\bA{{\bf A}}

\newcommand{\be}{\begin{equation}}
\newcommand{\ee}{\end{equation}}
\newcommand{\bea}{\begin{eqnarray}}
\newcommand{\eea}{\end{eqnarray}}

\newcommand{\p}{\partial}
\begin{document}

\begin{center}
\Large
{\bf Salpeter amplitudes  } \\
{\bf  in a Wilson-loop context } \\
\vskip 0.5cm
\large
 Ken Williams \\
{\small \em Continuous Electron Beam Accelerator Facility \\
        Newport News, VA 29606, USA \\
        and \\
\vspace*{-0.2cm}
Physics Department, Hampton University, Hampton, VA 29668}
\end{center}
\thispagestyle{empty}
\vskip 0.7cm

\begin{abstract}

\hspace{-8mm} The bound state problem for a gauge invariant quark-antiquark
system is considered in the instantaneous rest frame.
Focus here is on the long range non-perturbative interaction.
A two-time Green's function is constructed for Salpeter
amplitudes. The corresponding Schr\"{o}dinger equation is found to be in
the Salpeter form with a Wilson-loop term as the instantaneous kernel.
\end{abstract}

\newpage

\section{Introduction}

 The observed mass of a hadron results from the
 underlying dynamics of its constituent quarks and gluons.  Understanding the
 process is made difficult however by the complexities of low-energy QCD which
is widely believed to provide a mechanism for
 confinement.  This is one of the most interesting and difficult studies in
 quantum field theory.  The main problem stems from the essentially
 non-perturbative nature of confinement.  This is so except in certain cases.
For
 example, in the
 limit of heavy quark mass the theory is dynamically reduced to that of a
single particle, and at short distances asymptotic freedom suggests
 that the binding is dominated by one gluon exchange which in the static limit
 yields a coulombic interaction.  Thus heavy quarks forming a bound state
remain
 largely insensitive to the details of confinement, and their static properties
 are well described by non-relativistic constituent-quark potential models
 \cite{pot}.  Unfortunately, neither condition is met for the general case of
 deeply bound states, and a more fundamental approach is required.

\hspace{-8mm} The starting point for a relativistic description of the meson
spectrum is the
covariant Bethe-Salpeter(BS) equation.  It is the most orthodox
framework for addressing the two-body problem in quantum field theory and has
played a central role in the discussion for over three decades.  On the other
hand, the appearance of the relative time variable lacks a clear interpretation
in a Hamiltonian setting, and the usual approach is to make a three-dimensional
reduction of the kernel while retaining relativistic
kinematics.  In addition, one or both
single-particle propagators are often placed on an effective mass shell.
Symmetric treatment of this last
constraint leads to the Salpeter equation \cite{sal}
whose basic statement on the analytic form of its single-particle components is
given by $ \psi_\pm =  \Lambda^\pm \psi_\pm $ , also recognized as the
no-pair condition \cite{sucher}.  Apart from these issues are questions
about the Lorentz structure of the confining kernel.  Though the success
of phenomenological models suggest and lattice simulations confirm static
linear
confinement, going beyond this limit requires knowledge of the kernel's
dominant
Lorentz components.  But even this knowledge would shed no light on specifics
of
its dynamical origin.

\hspace{-8mm} What is needed and is essential for a discussion of
this problem in non-perturbative dynamics is a manifestly gauge invariant
formulation.  This is
   provided by the path integral method.  Defining the bound quark-antiquark
   amplitude in this way as a color singlet has led to some of the most
fruitful
   work in the non-perturbative regime beginning with the pioneering
   observations of Wilson \cite{wilson}.  The simple picture offered in his
area
   law for static quarks is compelling:  Contributions to the meson propagator
   fall off exponentially in the area swept out by lines of chromoelectric flux
   joining the constituents world lines; hence widely separated paths are
   suppressed.  The dominant effect is stated as an average over the transport
   of color along the contour of the minimum area,  $ \imath \log \langle
W({\cal
   C}) \rangle \approx \sigma S_{min} $.  The main difficulties here are
   technical and directly related to the fact that analytic expressions for the
   full single fermion propagator are not known.  For example, completely
static
   propagators lead to the well-known linear potential\cite{brown}, and leading
   corrections yield $ O(m^{-2}) $ spin-dependent\cite{eichten}, and more
   recently as given in the works of Brambilla and Prosperi\cite{bram},
   spin-independent contributions.  On another front, a full spinless
   propagator leads to relativistic flux-tube dynamics\cite{dubin}.

\hspace{-8mm} Here the following question is asked:  Can the Lorentz and
dynamical
uncertainties present in the Salpeter approach be resolved and the technical
difficulties of the path integral method overcome in a complimentary
formulation
which retains relativistic kinematics and spin structure?  The answer seems to
be
in the affirmative with a resulting Hamiltonian similar to that of a
relativistic flux-tube model\cite{olsson}.

\section{ Two-body propagator}

 The beginning parts of this section follow closely the treatment and notations
 found in ref\cite{bram} where the starting point is also the gauge invariant
 four-point Green's function
 \bea G(x_1,x_2,y_1,y_2) &=& \frac{1}{3}
 \langle 0| T \psi_2^c(x_2) U(x_2,x_1) \psi_1(x_1) \bar{\psi}_1(y_1)
 U(y_1,y_2)\bar{\psi}_2^c(y_2) |0 \rangle \\ &=& \frac{1}{3} \langle tr \,
 U(x_2,x_1) S_1(x_1,y_1 |A)U(y_1,y_2) C^{-1}S_2(y_2,x_2|A)C \rangle \,.
 \label{green}
 \eea
 The $ \psi $ 's here are fermion fields in the Heisenberg
 representation , $ U $ straight-line path ordered exponentials, and $ S $
single
 particle propagators in the external gauge field $ A $.  $ C $ is the charge
conjugation
 matrix while  $ c $ denotes
 charge conjugation.  The average is taken over gauge fields with virtual quark
loops ignored.
 Positive and negative frequency components of
 the full single particle propagator
 \bea
  \imath S(x,y|A) &=& \theta(x_0-y_0)
 S^+(x,y|A) -\theta(y_0-x_0) S^-(x,y|A) \label{g}
 \eea
  where $ S^+(x,y|A) =
 \langle 0| \psi^{(+)}(x) \bar{\psi}^{(-)}(y) | 0 \rangle $ and $ S^-(x,y|A) =
 \langle 0| \bar{\psi}^{(+)}(y) \psi^{(-)}(x) | 0 \rangle $ , satisfy the
 homogeneous Dirac equation
 \bea
  (\imath \not\hspace{-1.25mm}{\it D}_x - m) \,
 S^\pm (x,y|A) &=& 0 \label{homo}
 \eea
  with Cauchy condition
  \bea
  \left[
 S^+(x,y|A) + S^-(x,y|A) \right]_{x_0 =y_0} &=& \gamma_0 \delta^3(\bx -\by) .
 \label{cauchy}
 \eea
  In the instantaneous Salpeter approximation $
 \psi^{(\pm)} $ are expanded over free $\pm $ energy Dirac solutions,
 respectively, and thereby obey no-pair conditions \cite{sucher} which in the
 presence of the external field takes the form
 \bea
  \psi^{(\pm)} = \Lambda_\pm(\bpi ) \psi^{(\pm)} \label{np}
 \eea
 where $ \bpi =  - \imath \bnabla-g\bA $, and $ \Lambda_\pm(\bp) =
 (E_0 \pm H_0)/(2 E_0) $ are the usual energy projection operators with $ E_0 =
 (\bp^2+m^2)^{1/2} $ and $ H_0 = \gamma^0 (\bgamma \cdot \bp +m) $. The
boundary
 condition (\ref{cauchy}) is then modified to
 \bea
 S^{\pm}(x,y|A) \gamma^0 |_{x_0 = y_0}  &=& \Lambda_\pm( \bpi )
 \delta^3(\bx -\by)  \label{cau}
 \eea
 so that each term in (\ref{g}) independently has the property of a propagator.
 By standard
 methods \cite{lan} the solutions to (\ref{homo}) satisfying (\ref{cau})
 are expressed as the phase-space path integrals
 \bea
  S^\pm(x,y|A) &=&
 \int^\bx_\by {\cal D} [\bz,\bp] \, T_A \, T_s \; \Lambda_\pm(\bp -g\bA)
\label{prop2}
 \\ && \qquad \times \; \exp \left\{\imath \int^{x^0}_{y^0} dt [\bp \cdot
\dot{\bz}
 \mp E_0(\bp -g\bA) - gA_0 ] \right\} \gamma^0  \,. \nonumber
 \eea

\hspace{-8mm} In this expression $ T_A $ time orders
gauge-field operators for $ x_0 > y_0 $ and antitime orders
for $ x_0 < y_0 $ , while  $ T_s $ does the same for Dirac matrices. Terms
relevant
to the minimal-area relation enter upon a
semiclassical reduction. It is convenient to first
isolate Wilson-loop factors by translating the integration variable $ \bp \to
\bp + g\bA $.
The
momentum integration is then performed in the Gaussian approximation around
stationary paths $ \dot{\bz}_\pm = \pm \, \bp /E_0(\bp) $ yeilding
\be
S_{cl}^\pm(x,y|A) = \int^\bx_\by {\cal D} [\bz] T_A T_s \Lambda_\pm^\prime
\exp \left\{ \imath \int^{x^0}_{y^0} dt [ \mp m (1-{\dot{\bz}}^2)^{1/2}]
-\imath g \int_y^x A_\mu dr^\mu   \right\} \gamma^0  \label{cl}
\ee
The primed projector stands for evaluation at $ \bp_q \equiv \pm m
\dot{\bz}_\pm
(1-\dot{\bz}^2)^{-1/2} $ and an unimportant change in the integration measure
has been suppressed \cite{bram}.  Consider now the action of the four-gradient
\bea
\imath \p_\mu S^\pm_{cl} &=& {(p_\mu)}_{cl} S^\pm_{cl} \,.  \label{dcl}
\eea
Equation (\ref{homo}) with constraint (\ref{np}) follow from the quantization
of (\ref{dcl}) by the obvious transformations,  $ S^\pm_{cl} \to S^\pm
 $ , \, $ \bp_{cl}
\to - \imath \bnabla $ and $ \Lambda_\mp p_0 S^\pm \to 0 \,.  $ Equation
(\ref{cl}) along with the relation $ C^{-1} S^\mp(y,x|A) C = [- S^\pm(x,y| -
A^\tau)]^\tau $
, where $ \tau $ is the transpose operator, combine in the two-time Green's
function to
give
\bea
G(x_0;\bx_1,\bx_2|y_0;\by_1,\by_2) &=& \theta(x_0 - y_0) \, \hat{G}_+ +
\theta(y_0 - x_0) \, \hat{G}_-
\eea
 with definition
 \bea
 \lefteqn{
\hat{G}_\pm (x_0;\bx_1,\bx_2|y_0;\by_1,\by_2) = } \label{prop3} \\ &&
 \int^{\bx_1}_{\by_1} \int^{\bx_2}_{\by_2}  {\cal D}[\bz_1] \, {\cal D}[\bz_2]
 T_s \Lambda^{\prime (1)}_\pm \Lambda^{\prime (2)}_\pm \exp  \{ \imath
\int_{y_0}^{x_0} dt \sum_{j=1}^2 [ \mp m_j (1-{\dot{\bz}_j}^2)^{1/2} ] + \log
\langle W({\cal C}_\pm ) \rangle\} \gamma_1^0 \gamma_2^0 \nonumber
\eea
 for
the closed contours $ {\cal C}_\pm $

\begin{picture}(300,100)
\put(123,75){\vector(2,-1){5}}
\put(83,55){\vector(0,1){5}}
\put(83,80){\line(0,-1){50}}
\qbezier(83,80)(93,90)(123,75)
\qbezier(123,75)(143,65)(168,72)

\put(123,25){\vector(-2,1){5}}
\put(168,47){\vector(0,-1){5}}
\put(168,72){\line(0,-1){50}}
\qbezier(83,30)(93,40)(123,25)
\qbezier(123,25)(143,15)(168,22)

\put(265,75){\vector(-2,-1){5}}
\put(225,47){\vector(0,-1){5}}
 \put(225,72){\line(0,-1){50}}
 \qbezier(310,80)(285,90)(265,75)
 \qbezier(265,75)(235,65)(225,72)

 \put(265,25){\vector(2,1){5}}
 \put(310,55){\vector(0,1){5}}
\put(310,80){\line(0,-1){50}}
\qbezier(310,30)(285,40)(265,25)
\qbezier(265,25)(235,15)(225,22)

 \put(76,90){y1}
 \put(78,20){y2}
 \put(163,82){x1}
 \put(163,12){x2}
\put(220,82){x1}
\put(220,12){x2}
\put(305,90){y1}
\put(305,20){y2}
\put(123,51){+}
\put(265,51){--}

\put(185,0){fig(1)}

\end{picture}
\\
time increasing to the right. Here the compatibility of the instantaneous
Salpeter picture with that of the
Wilson loop is apparent: In addition to the intermediate
propagation of two particles, both allow for the propagation of two
antiparticles
(double Z-graphs in time-ordered perturbation theory ) evidently represented
here by $ \hat{G}_- $. Single Z-graph
components do not enter. As mentioned earlier, the
straight-line approximation to the minimal-area law, $ \imath \log \langle
W({\cal C_\pm}) \rangle\ = \sigma S_{min} \equiv   \int^{t_>} dt L_g = \pm
\int^{x_0} dt L_g $ is taken. In the center-of-momentum
frame
\bea
L_g &=& \sigma |\br| \int_0^1 ds \gamma^{-1}(\bv_t)
\eea
 where $ \gamma(\bv) =
(1-\bv^2)^{-1/2} $ and $ \bv_t = s \dot{\bx}_{1\perp} + (1-s)
\dot{\bx}_{2\perp}
$ with $ v_{i\perp} \equiv (\delta_{ij} - \hat{r}_i \hat{r}_j ) v_j $ as the
i-th component of $ \bv $ perpendicular to $ \br = \bx_1 -\bx_2 $. With $ L_q
\equiv  \sum_{j=1}^2  m_j (1-{\dot{\bz}_j}^2)^{1/2} $  the exponential argument
in (\ref{prop3}) can be written
\be
\imath {\cal A}_\pm = \mp \imath \int^{x_0}_{y_0} dt ( L_q + L_g) = \mp \imath
\int^{x_0}_{y_0} dt L .
\ee
{}From the stationary condition, action of the four-gradient yields
\bea
 \sum_j -\imath \bpartial_j (\imath {\cal A}_\pm ) &=& \sum_j -\imath
\frac{d}{dx_o} \bpartial_{\dot{\bx}_{j \pm}} (\imath {\cal A}_\pm ) \\
&=& \sum_j \pm m_j \dot{\bx}_{j \pm} \gamma( \dot{\bx}_j ) + \sigma |\br|
\int^1_0 ds \gamma(\bv_t) \bv_t \\ &\equiv & \sum_j \bp_{qj} +\bp_g
\eea
and
\bea
\imath \partial_{x_0}   (\imath {\cal A}_\pm ) &=& \pm (\sum_j  \dot{\bx}_{j
\pm} \cdot \frac{\p L}{\p \dot{\bx}_{j \pm}} - L ) \\ &=& \pm (\sum_j m_j
\gamma (\dot{\bx}_j ) + \sigma |\br| \int^1_0 ds \gamma(\bv_t) ) \\ &\equiv &
\pm (\sum_j m_j \gamma (\dot{\bx}_j ) + V ) \, .
\eea
Then
 \bea
\sum_j -\imath \bpartial_j \, \hat{G}_\pm &\to & ( \sum_j \; \bp_{qj} + \bp_g)
\, \hat{G}_\pm  \label{mom}
\\ \imath \p_{x_0} \, \hat{G}_\pm &\to & (
H_{01}^\prime + H_{02}^\prime \pm V_\pm \, ) \, \, \hat{G}_\pm
\eea
with $
V_\pm = \Lambda^{\prime (1)}_\pm \Lambda^{\prime (2)}_\pm \, V \,
\Lambda^{\prime (1)}_\pm \Lambda^{\prime (2)}_\pm $.

\section{ Eigenvalue equation }

Given $ \chi^{(+)}(t;\br_1,\br_2 ) $ and $ \chi^{(-)}(t^{\prime};\br_1,\br_2 )
$
 as positive and negative frequency components of the single-time bound state $
 q\bar{q} $ amplitude with $ t < t^{\prime} $, the freely propagating amplitude
 at intermediate times is found with help from the Green's function
 \bea
 \chi(x_0;\bx_1,\bx_2) &=& \imath \int d^3 r_1 d^3 r_2 [
 G(x_0;\bx_1,\bx_2|t;\br_1,\br_2) \chi^{(+)}(t;\br_1,\br_2 ) \\ && \qquad
\qquad
 \qquad \qquad - G(x_0;\bx_1,\bx_2|t^{\prime};\br_1,\br_2)
 \chi^{(-)}(t^{\prime};\br_1,\br_2 ) ] \nonumber
 \eea
 and so satisfies the
 time-independent Schrodinger equation
 \bea
 \imath \p_{x_0} \, \chi &=& H \,
 \chi
 \eea
 where the Hamiltonian is given by
 \bea
 H &=& H_{01}^\prime +
 H_{02}^\prime + V_+ - V_- \,.  \label{ham}
 \eea
 Here $ H_0^\prime $
 represents the quark's kinetic energy and $ V $ the gluon field
 contribution from the Wilson-loop (which apparently enters as the
instantaneous
 BS kernel). Projection operators on $ V $ prevent mixing with the
 non-normalizable continuum states of the same energy.

\hspace{-8mm} This result should be compared with others from the Wilson-loop
formalism.  The classical limit to
the spinless Hamiltonian of ref\cite{dubin} is recovered from (\ref{ham}) by
the
obvious reduction, $ H_0 \to + E_0 $, which incorporates the no-backtracking
constraint; that is, the second loop of fig(1) does not contribute ( $ V_- \to
0 $) in this approximation.  $ O(1/m^2) $ terms of $ (H)_{12} $ in the standard
Dirac representation
give leading relativistic corrections to the static $ \sigma r $ interaction;
the reduction is equivalent to a Foldy-Wouthuysen transformation. In the center
of momentum frame the corrections are
\bea
V_{corr} & \simeq & - \sigma (1/m_1^2 +1/m_2^2 - 1/(m_1 m_2 ) ) L^2 /(6r)
\label{nr} \\ && \qquad \qquad \qquad \qquad - \sigma (\bL \cdot \bs_1 /m_1^2 +
\bL \cdot \bs_2 /m_2^2 ) /(6r) \,.  \nonumber
\eea
 The spin-independent term
above agrees with the semirelativistic result of ref\cite{bram} omitting a
angular-momentum independent contribution resulting from their particular
operator
ordering prescription, $ V_{cl}(\br,\bp) \to \{ V_{cl} \}_{ord}
\equiv V_{qm} $. An alternative prescription better suited to quantization of
(\ref{ham}) would be in terms of classical coordinate and velocity observables,
$ (\br,\bv) $. In this
case the $ \bv_\perp $ operators are found as truncated matrices
in a suitable basis from the symmetrized orbital angular momentum equation, $
\bL(\br,\bv_\perp) = \bx_i \times \bp_i $, for a given state.  The energy
eigenvalue
equation thus becomes a matrix equation in the radial coordinate only and is
solved
variationally.  For details of the method see e.g. ref\cite{colin}.

\hspace{-8mm} Spin corrections responsible for fine and hyperfine structure
of the spectrum are of course an important and delicate concern.
They must be handled carefully.  The numerical coefficient of the spin-orbit
term
in (\ref{nr}), -1/6, is at variance with the -1/2 factor found in most of the
literature.  Most often the -1/2 follows from an assumption of
dominant scalar confinement\cite{scalar} which in relativized models based on a
BS reduction is thought to be spectroscopically favored over other Lorentz
forms.  On the
other hand, this same disagreement appears with the $ O(1/m^2) $ Wilson-loop
result of
ref\cite{bram} where no such assumption is made. From a purely
mathematical point of view this results from the different ways in which the
holonomic straight-line condition is applied to the minimal-area
relation; a factor of
-1/6 (-1/2) results when the constraint is imposed before (after) the variation
implied here in equation(\ref{mom}) and in appendix B of ref\cite{bram}.
Only the former option, followed here, is in line with proper procedures for
the
mechanics of constrained systems. It should be pointed out that Gromes has
deduced a
relation\cite{gromes} in support of the -1/2 factor (and scalar confinement)
 from arguments of Lorentz covariance following a reasoning in Representation
 Theory on
the group structure of Poincar\'{e} transformations. The relevance of this
formalism in the present rest-frame Hamiltonian context is however not entirely
clear and the question will not be entered into here.

\section{Summary}

    Salpeter amplitudes have been introduced into the path integral formulation
of QCD
beginning from the gauge-invariant four-point function.  A
Hamiltonian description for mesons as bound quark-antiquark states has been
derived in a systematic and straightforward manner.  Interestingly, a time
derivative of the Wilson-loop operator appears as the instantaneous BS kernel.
Both
relativistic kinematics and spinor structure of the amplitude have been
preserved.  The focus here has been on long range non-perturbative effects of
the gluon dynamics, though a realistic calculation of the spectrum must
take medium and short range contributions into account as well.  These
enter asymptotically as a coulombic interaction which is easily added to the
present result.
An apparent discrepancy between the $ O(1/m^2) $ spin-orbit coefficient and the
relation of
Gromes has been noted and is under study.

\vspace{10mm}

\hspace{-8mm} {\sl Acknowledgments}: This work was supported
in part by the National Science Foundation under Grant No. HRD-9154080.

\end{document}